\documentclass[pre]{revtex4}
\usepackage{graphicx}
\usepackage{amsmath}

\newcommand{\PP}{{\mathcal{P}}}

\newcommand{\RR}{{\bf R}}

\newcommand{\kk}{\mathbf{k}}
\newcommand{\rr}{{\bf r}}

\newcommand{\dd}{{\bf d}}

\newcommand{\BE}{\begin{equation}}
\newcommand{\EE}{\end{equation}}
\newcommand{\BEN}{\begin{eqnarray}}
\newcommand{\EEN}{\end{eqnarray}}

\newcommand{\LL}{{\mathbf L}}

\newcommand{\EW}{{\rm EW}}

\newcommand{\beq}{\begin{equation}}
\newcommand{\eeq}{\end{equation}}
\newcommand{\beqs}{\begin{eqnarray}}
\newcommand{\eeqs}{\end{eqnarray}}

\begin{document}

\title{Computation of the High Temperature Coulomb Density Matrix\\in Periodic Boundary Conditions} 

\author{B. Militzer}
\address{Department of Earth and Planetary Science and Department of Astronomy,
         University of California, Berkeley, CA 94720}

\date{\today}

\begin{abstract}
      The high temperature many-body density matrix is fundamental to
      path integral computation. The pair approximation, where the
      interaction part is written as a product of pair density
      matrices, is commonly used and is accurate to order $\tau^2$,
      where $\tau$ is the step size in the imaginary time. Here we
      present a method for systems with Coulomb interactions in
      periodic boundary conditions that consistently treats the all
      interactions with the same level of accuracy. It shown that this
      leads to a more accurate high temperature solution of the Bloch
      equation. The method is applied to many-body simulation and
      tests for the isolated hydrogen atom and molecule are presented.
\end{abstract}

\maketitle

\section{Introduction}

Quantum Monte Carlo (QMC) methods are frequently used to study
interacting many-body systems in different fields of physics and
chemistry when a high degree of accuracy is needed~\cite{FM99}. The
description of correlation effects combined with favorable scaling
properties of $N^3$ or better (N is the number of particles) make
these techniques effective in many applications. Path integral Monte
Carlo (PIMC) is unique among other QMC methods because it can describe
quantum systems at {\em finite
temperature}~\cite{Fe53,Fe72,PC84,PC87,Ce95}. The method is based on
the thermal density matrix that characterizes the properties of a
system in thermal equilibrium.

Many applications of the PIMC method require an accurate treatment of
Coulomb interactions in periodic boundary conditions (PBC). This
includes all electronic structure simulations that describe electrons
as individual particles. However, there is still no universally
accepted method to compute the Coulomb propagator in PBC, which has
led to unnecessary approximations resulting in less efficient and less
accurate many-body simulations.

In this article, we describe an accurate approach to compute the
Coulomb pair density matrix in a periodic system. It can easily be
generalized to other long range interactions~\cite{MG06}. 
The density matrix of a bosonic (B) or fermionic (F) system at
temperature $T$ can expressed in terms of an imaginary-time path
integral,
\beq
\label{PI}
\rho_{\rm B/F}(\RR,\RR';\beta)=\frac{1}{N!} \sum_\PP \: (\pm 1)^\PP 
\! \! \! \! \! \!  \int\limits_{\RR \rightarrow \PP\RR'} \! \! \! \! \! \! d\RR_i \:
\rho(\RR,\RR_1;\tau)\,\rho(\RR_1,\RR_2;\tau) \ldots \rho(\RR_{M-1},\PP \RR';\tau)
\;,
\eeq
where $\tau$ is the time step $\tau=\beta/M$ with $\beta=1/k_{\rm B}
T$, and $k_{\rm B}$ is Boltzmann's constant. $N$ particles in
real-space representation, $\RR=\{\rr_1,\ldots, \rr_N\}$, are described in
the canonical ensemble. Atomic units of Bohr radii and Hartree will be
used throughout this work.

Instead of requiring the propagator at temperature, $T$, path
integrals rely a density matrix at much higher temperature, $M \times
T$. At high temperature, the many-body density matrix can be
computed with good accuracy because exchange effects as well as
three-body correlations are negligible in the limit of high
temperature. A novel method of constructing the high temperature
density matrix (HTDM) for systems with {\em long-range interactions}
with {\em periodic boundary conditions} is the focus of this study. 

Different approximations for the Coulomb HTDM have been
advanced that all become exact in the limit of $\tau \to 0$. However,
all PIMC simulations are performed at finite $\tau$. Therefore an
accurate representation of the HTDM is very important. Its accuracy
determines the maximum time step $\tau$ one can use. A larger time
step allows one to significantly cut down on the number of time slices
in the path integral, $M$, and therefore improve the efficiency of many-body
simulations. This gain in efficiency may be crucial in practical
applications to perform accurate simulations at low temperature or for
large systems.

The HTDM can be used to define the potential action,
$U(\RR,\RR';\tau)$. For a system of $N$ distinguishable particles in
real space representation this reads,
\beq
\rho(\RR,\RR';\tau)=\exp \left\{-U(\RR,\RR';\tau) \right\} \,\prod_{i=1}^N \rho_0(\rr_i,\rr_i';\tau) 
\;\;.
\eeq
$\rho_0$ is the free particle density matrix in $D$ dimensions,
\beq
\rho_0(\rr,\rr';\tau) = (4 \pi \lambda \tau)^{-D/2} \exp \{ -(\rr-\rr')^2/4\lambda\tau\}
\;\;.
\eeq
$\lambda$ is the a mass dependent parameter, $\hbar^2/2m$. $U$ is commonly approximated
as the product over the nonideal parts of all pair density matrices $\rho(\rr_{ij},\rr_{ij}';\tau)$,
\beq
\exp \left\{-U(\RR,\RR';\tau) \right\} \approx
\exp \left\{-\sum_{i<j}u(\rr_{ij},\rr_{ij}';\tau) \right\} 
= \prod_{i<j} \frac{\rho(\rr_{ij},\rr'_{ij};\tau)}
{\rho_0(\rr_{ij},\rr'_{ij};\tau)}
% We cannot write the product because of the center-of-mass term
% {\rho_0(\rr_i,\rr_i';\tau) \rho_0(\rr_j,\rr_j';\tau)}
\;\;.
\label{action}
\eeq
This is called the {\em pair approximation}, and is accurate to order
$\tau^{2}$~\cite{PC84,PC87,Ce95}. One essentially starts with a exact
solution of the two-body problem when performing many-body
simulations. This mean only 1 time slice is needed to study the
hydrogen atom at any temperature and equilibrium ionization state but
more are needed for the hydrogen molecule when three-body correlations
are important. Exchange effects have been neglected in
Eq.~\ref{action}, which justified for a small time step $\tau$. In the
full path integral, they enter as a sum of over permutations in
Eq.~\ref{PI}.

For a Coulombic system with periodic boundary conditions, these pair
density matrices, $\rho(\rr,\rr';\tau)$, are solutions of the two
particle Bloch equation with the Ewald potential~\cite{Ew17},
\beq
\frac{\partial \rho}{\partial \tau} \,=\, -\hat{H} \rho 
\,=\, \lambda_{ij} \, \nabla^2_\rr \, \rho \,-\, V_{\rm EW}(\rr) \rho
\;,
\label{Bloch_eq}
\eeq
with the initial condition $\rho(\rr,\rr';\tau=0) =
\delta(\rr-\rr')$. The reduced mass $\mu_{ij}=m_i m_j/(m_i+m_j)$ enters
through $\lambda_{ij} \equiv \hbar^2/2\mu_{ij}$. The pair density matrix
can also be derived from the Feynman-Kac relation,
\beq
\rho(\rr,\rr';\tau) = \rho_0(\rr,\rr';\tau) \; 
\left< e^{- \int_0^\beta dt \: V (\rr(t)) } \right>_{\rr \rightarrow \rr'}
\;,
\label{FK}
\eeq
where the average is to be taken over all free-particle (Brownian)
paths from $\rr$ to $\rr'$. For potential without negative
singularities, this expression can easily be evaluated numerically for
specific pairs of $\rr$ and $\rr'$ in order to verify the
approximations to be discussed below. However, the results will always
have a statistical uncertainty due to the finite sample of Brownian
paths.

The resulting Ewald pair action, $u_{\rm EW}(\rr,\rr';\tau) = - \ln
[\rho_{\rm EW}/\rho_0)]$, determines the weight of the paths
in~Eq.~\ref{PI} where $\rr$ and $\rr'$ represent the separations of
pairs of particles $i$ and $j$ at two adjacent time slices,
\beq
\rr = \rr_i(t) - \rr_j(t) \quad,\quad \rr' = \rr_i(t+\tau) - \rr_j(t+\tau) \;\;.
\eeq
For a spherically symmetric potential, the pair density matrix depends
on $\tau$ and three spatial variables: the initial and final pair
separations $|\rr|$ and $|\rr'|$ as well as the angle between them
$\theta$. Alternatively, it can be expressed in terms of the variables
$q$, $s$ and $z$,
\beq
q \equiv\frac{1}{2}(|\rr|+|\rr'|) \;,\; 
s \equiv |\rr-\rr'| \;,\; 
z \equiv |\rr|-|\rr'|.
\eeq
For Coulomb potential, the dependence on $z$ drops
out~\cite{Ho63}. The Ewald potential, however, has the symmetry of the
periodic simulation cell which requires both the initial and final
pair separation to be specified with respect to the cell. This implies
that the pair density matrix for the Ewald potential depends on
six spatial variables. This makes computation and storage of the
corresponding action extremely awkward. 

Previous methods~\cite{PC94,Ma96} to deal with this difficulty have
involved a break-up of the Ewald potential into a spherically
symmetric short-range piece and a long-range remainder, 
\beq
V_{\rm EW}(\rr) = V_{\rm s.r.}(|\rr|) + V_{\rm l.r.}(\rr)\;. 
\label{breakup}
\eeq 
The short-range piece has been treated numerically using the matrix
squaring technique developed by Storer~\cite{St68}. In principle, it
allows one to derive the exact action for a spherically symmetric
potential but in practice the accuracy is controlled by numerical
accuracy of the integration. Matrix squaring is performed on a grid
and controlling the associated grid errors requires significant
care~\cite{KE06}. To treat the cusp condition at the origin
accurately, the short-range part must include the singular part of the
potential. Most simply, one can use the direct $1/r$ interaction term
as short-range part. Alternatively, one can employ the optimized Ewald
break-up method described in~\cite{Na95}, which allows one to
construct a short-range piece that always decays within the boundaries
of the simulation cell. A detailed review of the break-up method and
its accuracy is given in~\cite{KE06}. The {\em primitive
approximation},
\beq
u_{\rm P.A.}(\rr,\rr';\tau) = \frac{\tau}{2} \left[ V(\rr)+V(\rr') \right]\;,
\label{PA}
\eeq
provides a simple straightforward way to add the long-range remainder,
\beq
u(\rr,\rr';\tau) \approx u_{\rm s.r.}(\rr,\rr';\tau) + 
\frac{\tau}{2} \left[ V_{\rm l.r.}(\rr)+V_{\rm l.r.}(\rr') \right]\;.
\label{prim-approx}
\eeq
but the random phase approximation has also been applied to the
long-range action~\cite{Ma96}. When we later refer to
Eqs.~\ref{breakup} and \ref{prim-approx}, we assume that the $1/r$
potential has been used as short-range piece.

The break-up of the Ewald potential~\cite{PC94,Ma96} introduces an
additional approximation to path integral computation. While it has
been successfully in many applications~\cite{Jo96,MC00,MC01,Zo02}
there is a need for improvement. The break-up is {\em ad hoc} and
introduces some arbitrariness to the construction of the Ewald
action. The accuracy remains high if either potential piece is
sufficiently smooth on the scale of the thermal de Broglie wave
length, $\lambda_d = \sqrt{4 \pi
\lambda \tau}$. However, the convergence to the correct answer as
function of cell size and temperature is difficult to assess. For
these reasons, we present a method here that consistently treats the
direct Coulomb interaction within the simulation cell and that with
periodic images with the same level of accuracy. Our method thereby
avoids introducing a short and long-range potential. The necessary
pair density matrices are derived for the Coulomb potential, which
makes it possible to utilize the large amount of analytic and
numerical work available for this potential. The development of the
method proceeds by three steps which are now described.

\section{Method}

\subsection{Computation of the pair density matrix for the Coulomb potential}

It is first necessary to compute the pair density matrix for an
isolated pair of particles. This can either be done with matrix
squaring or from the sum over eigenstates. In the matrix squaring
technique, one can take advantage of the fact that only s-wave
contributions enter in case of the Coulomb potential as done by
Storer~\cite{St68}. In the squaring technique, one starts from a high
temperature expansion and then numerically squares the density matrix
using,
\beq
\rho(\rr,\rr';2\tau) = \int \dd\rr'' \, \rho(\rr,\rr'';\tau) \, \rho(\rr'',\rr';\tau)\;,
\eeq
until the lowest temperature has been reached. This typically requires
10--30 iterations. Since matrix squaring is done on a grid in $\rr$
and $\rr'$ a grid error is introduced. For the Coulomb potential this
usually requires a high number grid points near the singularity at the
origin. This singularity also requires care when the density matrix is
initialized using a high temperature expansion at the beginning of the
squaring procedure. Despite these shortcomings, the squaring technique
has the advantage that it can easily be applied to arbitrary spherical
potentials~\cite{KS73}. Alternatively, Schmidt and Lee developed an
approach where kinetic and potential operators are applied repeatedly
using Fourier transforms~\cite{KS95}. This method has been adopted by
J. Shumway to study a variety of systems with
PIMC~\cite{Sh06}. Vieillefosse used an analytic power-series expansion
method to construct the action~\cite{Vi94}.

For potentials that do not exhibit negative singularities, the pair
density matrix can also be derived from the Feynman-Kac formula,
Eq.~\ref{FK}. This approach was applied to the Yukawa interaction in
Ref.~\cite{MG06}. It avoids introducing grid errors and readily yields
diagonal matrix elements as well as the first term in an expansion
for off-diagonal elements but it impractical for
constructing a table the includes arbitrary off-diagonal
elements. Matrix squaring would then be more appropriate~\cite{KE06}.

In this paper, we rely on another approach and compute the pair
density matrix by summation over eigenstates~\cite{Po88},
\beq
\rho(\rr,\rr';\tau) = \sum_s e^{-\tau \epsilon_s} \, \psi_s(\rr) \, \psi_s^*(\rr')\;. 
\eeq
It also avoids grid errors and can provide off-diagonal elements
efficiently but requires that the eigenstates are known with high
precision, which is the case for the Coulomb potential. Although the
number of states to be considered increases with
temperature but the summation can be performed accurately for the
temperatures of interest as was shown in Ref.~\cite{Po88}.

\begin{figure}[htb]
\centerline{\includegraphics[angle=0,width=0.85\textwidth]{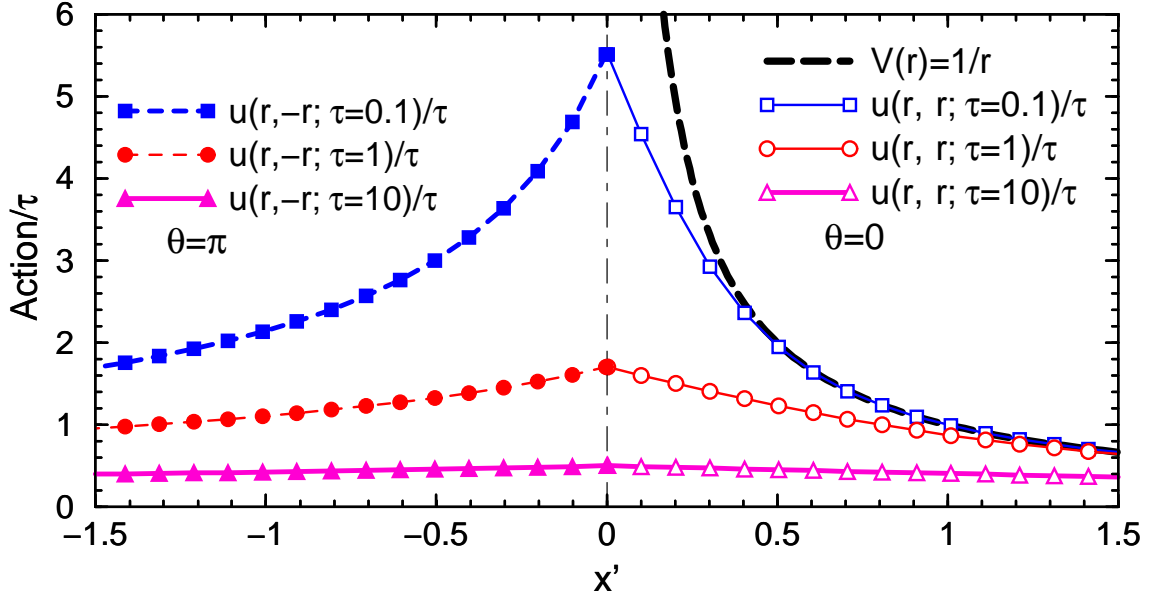}}
\caption{Potential action divided by the inverse
         temperature $\tau$ at $\tau=0.1,1.0,10.0$ for an isolated
         pair of electrons as a function of $x'$ with $\rr=(|x'|,0,0)$
         and $\rr'=(x',0,0)$. With increasing separation, the diagonal
         action $(\rr=\rr')$ converges rapidly to the primitive
         action, $\tau/r$. The action in exchange case $(\rr=-\rr')$
         is consistently higher than the diagonal action,
         $u(\rr,-\rr,\tau) \ge u(\rr,\rr,\tau)$.}
\label{ee-action}
\end{figure}

For illustration, Fig.~\ref{ee-action} shows examples of the electron
pair potential action for the on diagonal case ($\rr'=\rr$) and the
exchange case ($\rr'=-\rr)$ over two decades of $\tau$. The cusp
condition leads to a linear behavior near the origin. For large $r$,
the diagonal action converges to the primitive action. The exchange
term always has a higher action, which can be understood from the
Feynman-Kac formula in Eq.~\ref{FK} where one averages over all
Brownian paths. While in the exchange case paths beginning at $\rr$
must diffuse around the repulsive $1/r$ potential to reach $-\rr$, in
the diagonal case can avoid region of high potential more easily. 

%%%%%%%%%%%%%%%%%%%%%%%%%%%%%%%%%%%%%%%%%%%%%%%%%%%%%%%%%%%%%%%%%%%%%%%%%%%%%%%%%%%%%%%%%%%%%%%%%%%%%%%%%%%
%%%%%%%%%%%%%%%%%%%%%%%%%%%%%%%%%%%%%%%%%%%%%%%%%%%%%%%%%%%%%%%%%%%%%%%%%%%%%%%%%%%%%%%%%%%%%%%%%%%%%%%%%%%
%%%%%%%%%%%%%%%%%%%%%%%%%%%%%%%%%%%%%%%%%%%%%%%%%%%%%%%%%%%%%%%%%%%%%%%%%%%%%%%%%%%%%%%%%%%%%%%%%%%%%%%%%%%

\begin{figure}[t]
\centerline{\includegraphics[angle=0,width=0.85\textwidth]{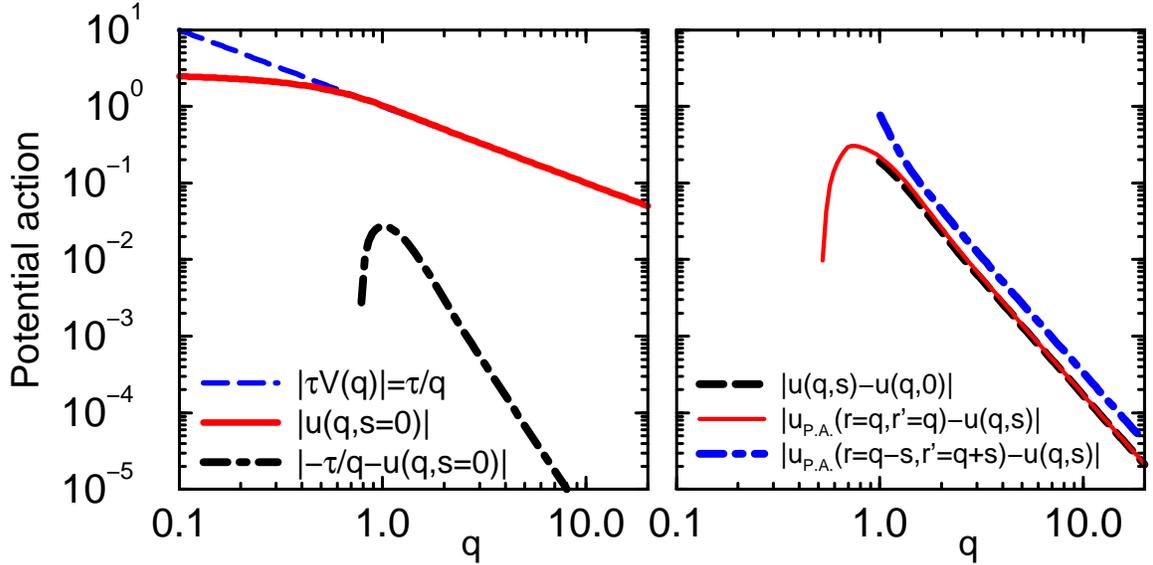}}
\caption{The potential pair action $u(q,s)$, shown for a proton-electron 
       pair at $\tau=1$ shown as function of
       $q=\frac{1}{2}[|\rr|+|\rr'|]$. The left graphs shows that, for
       {\em diagonal} configurations $\rr=\rr'$, the error in the
       primitive approximation, $-\tau/q$, decays like $q^{-4.1}$. The
       right graphs shows that the {\em off-diagonal} terms decays
       like $q^{-3.1}$ for fixed $s=|\rr-\rr'|=2 \sqrt{\lambda\tau}=
       \sqrt{2}$ (thick dashed line). The dot-dashed line shows the
       error in the primitive approximation for parallel $\rr$ and
       $\rr'$ ($\theta=0$). The solid lines shows this error in the
       case of $|\rr|=|\rr'|$ and angle $\theta>0$.  }
\label{offdiag}
\end{figure}

\begin{figure}[htb]
\centerline{\includegraphics[angle=0,width=0.5\textwidth]{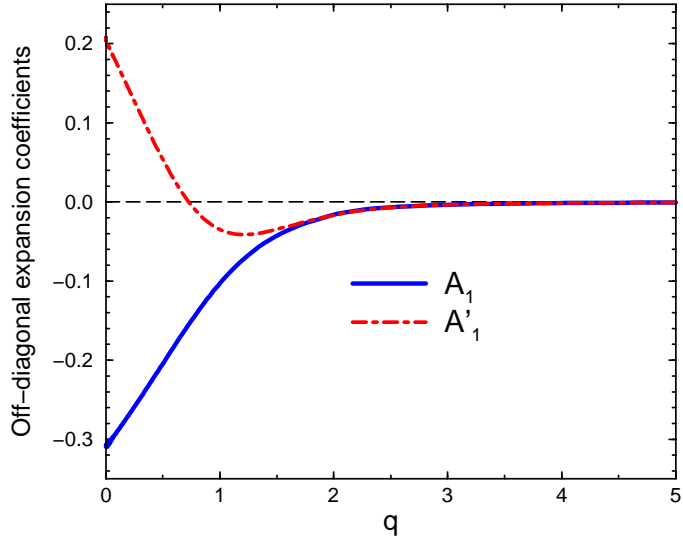}}
\caption{Off-diagonal expansion coefficients for proton-electron 
	 pair action $A_1$ and its $\tau$ derivative $A'_1$ at $\tau=1$.}
\label{offd-exp}
\end{figure}

To increase the efficiency of PIMC simulations, the HTDM is derived
beforehand and tabulated. For a given time step $\tau$, the Coulomb pair
density matrix is a function of the two variable $q$ and $s$, $u(q,s)
\equiv u(\rr,\rr';\tau)$. In order to reduce the storage, we fit the
off-diagonal terms using an expansion of powers of $s^2$ (compare
with~\cite{Ce95}),
\beq
u(q,s) \approx 
u(q,0) + \sum_{i=1}^{n} A_{i}(q) \, s^{2i},
\label{offd-s}
\eeq
For each $q$, a least squares fit can represent the off-diagonal term
with sufficient accuracy. The results for $n=1$ are shown in
Fig.~\ref{offd-exp}. The expansion works well in PIMC because the
kinetic energy prevents adjacent point on the path to exceed
separation much larger than the thermal de Broglie wave length. The
storage of complete set of off-diagonal terms is therefore not needed.
For the Coulomb potential, the off-diagonal terms decay approximately
like $q^{-3.1}$, which is demonstrated by three curves in
Fig.~\ref{offdiag}. The first compares the diagonal ($s=0$) and
off-diagonal action ($|r|=|r'|=q$, $s=2 \sqrt{\lambda\tau}=\sqrt{2}$)
at constant $q$. The magnitude of $s$ represents a typical separation
of adjacent points along the path which is mostly determined by the
free particle action. The second curve in Fig.~\ref{offdiag} shows the
error in the primitive approximation for the same $q$ and $s$
parameters. The third curve shows the error in the primitive
approximation for parallel vectors ($\theta=0$) with the exact
action. The graph underlines that the sum over off-diagonal
contributions from the image charge converges. Also for large
separations, the primitive approximation works well, and off-diagonal
contributions may be neglected.

These properties is used in PIMC simulations, where one typically
computes the off-diagonal contributions only for particle pairs within
the simulation cell using the minimum imagine convention. To date, no
efficient way to sum up all off-diagonal contributions has been
proposed. However, if necessary, a summation over the required number
of images can easily be included.

%%%%%%%%%%%%%%%%%%%%%%%%%%%%%%%%%%%%%%%%%%%%%%%%%%%%%%%%%%%%%%%%%%%%%%%%%%%%%%%%%%%%%%%%%%%%%%%%%%%%%%%%%%%
%%%%%%%%%%%%%%%%%%%%%%%%%%%%%%%%%%%%%%%%%%%%%%%%%%%%%%%%%%%%%%%%%%%%%%%%%%%%%%%%%%%%%%%%%%%%%%%%%%%%%%%%%%%
%%%%%%%%%%%%%%%%%%%%%%%%%%%%%%%%%%%%%%%%%%%%%%%%%%%%%%%%%%%%%%%%%%%%%%%%%%%%%%%%%%%%%%%%%%%%%%%%%%%%%%%%%%%

\subsection{Pair approximation for the image charges}

The pair approximation Eq.~\ref{action} is commonly used to
approximate the action of a path. This expression follows from the
Feynman-Kac formula Eq.~\ref{FK}, here written for a {\em finite} system of
$N$ particles, 
\beqs
\exp \left\{ -U(\RR,\RR';\tau) \right\} 
 &=&\left< \exp\left\{-\int_{0}^{\tau} \!\!\!\!  dt \sum_{i<j}^{N}  \frac{Q_iQ_j}{|\rr_{ij}(t)| }\right\}
       \right>_{\RR \rightarrow \RR'}\\
 &=&\left< \prod_{i<j}^N \exp\left\{-\int_{0}^{\tau}\!\!\!\! dt \; \frac{Q_iQ_j}{|\rr_{ij}(t)|} \right\}
       \right>_{\RR \rightarrow \RR'}\\
 &\approx& \prod_{i<j}\left<\exp\left\{-\int_{0}^{\tau}\!\!\!\!dt \; \frac{Q_iQ_j}{|\rr_{ij}(t)|}
          \right\} \right>_{\rr_{ij} \rightarrow \rr'_{ij}}    \label{line4} \\
 &=& \prod_{i<j}\exp\left\{-u(\rr_{ij},\rr'_{ij};\tau) \right\} \\
 &=& \exp\left\{-\sum_{i<j} u(\rr_{ij},\rr'_{ij};\tau) \right\}
    \label{eq1}
\eeqs
The pair approximation enters in~Eq.~\ref{line4} where it is assumed
that interactions between pairs of particles may be averaged
independently from the location of paths of the remaining
particles. This approximation becomes increasingly accurate for large
separations. It is exact at high temperature and, as stated before, the
error is of $O(\tau^{3})$.

Now we will describe how the pair approximation can be applied to the
systems with periodic boundary conditions. In principle, one only has
to add sum over the interaction with of all periodic images using the
pair approximation, which assumes that the diffusion of paths of the
image particles may also be averaged independently. Since the error in
the pair approximation also decreases with separation, applying it to
the periodic images is less of an approximation than one is using
already by applying the pair approximation to the direct interaction
within the simulation cell.

However, Eq.~\ref{eq1} converges {\em conditionally} since the Coulomb
potential is long-ranged. This problem has been solved by introducing
the Ewald potential which assumes charge neutrality guaranteed by a
neutralizing background. Starting from the Ewald potential, we will
now give an approximate expression for the corresponding Ewald
action. Since the exact action converges to the primitive
approximation for large separations (Fig.~\ref{offdiag}), the
quantum correction is a short-ranged function,
\beqs
\Delta u(\rr,\rr';\tau) = u(\rr,\rr';\tau) 
                        - \frac{\tau}{2} Q_i Q_j \left(\frac{1}{|\rr|}+\frac{1}{|\rr'|} \right)\;.
\eeqs
Since it decays faster than $r^{-3}$ (Fig.~\ref{offdiag}), the sum
over all images converges. It should be noted that $u(\rr,\rr';\tau)$
does not need to go to zero with the simulation cell. This allows us
to construct an Ewald action using the primitive approximation and a
sum of quantum corrections $\Delta u$,
\beq
\label{u_EW}
u_\EW(\rr,\rr';\tau) 
\approx \frac{\tau}{2} Q_i Q_j [V_\EW(\rr)+V_\EW(\rr')] 
+ \sum_{\LL} \Delta u(\rr+\LL,\rr'+\LL;\tau) 
+ u_{\rm BG}\;\;,
\eeq
which is the central result of this article. We will discuss diagonal
matrix elements first, off-diagonal elements are derived in the next
section. Note that it is periodic as well as symmetric in the
arguments $\rr$ and $\rr'$, and satisfies the cusp condition, i.e. the
$1/r$ singularity in the Coulomb potential as $r\rightarrow 0$ is
canceled in the Bloch equation by a term arising from the kinetic
energy operator.

\begin{figure}[htb]
\centerline{\includegraphics[angle=0,width=0.85\textwidth]{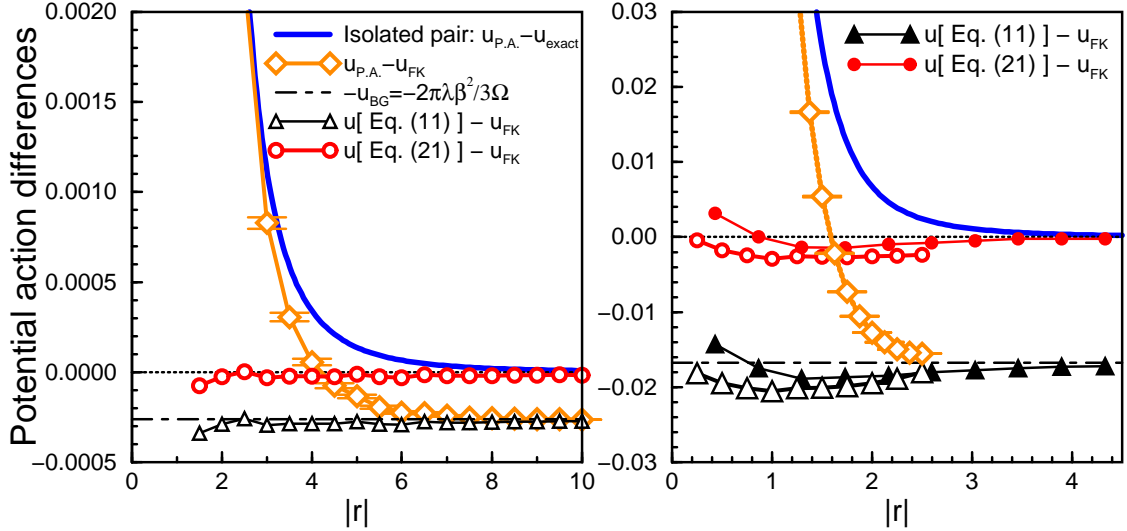}}
\caption{Various approximations for the Ewald action for pair of electrons 
         at $\tau=1$ are compared on the diagonal as a function of
         $|\rr|$ with the exact action $u$, which was either computed
         analytically or using the Feynman-Kac formula. The left graph
         shows action difference along the (1,0,0) direction in a
         cubic cell of 20 a.u. size. The right graph repeats all those
         curves with open symbols for a smaller cell of 5 a.u. but
         results for the (1,1,1) direction have been added with filled
         symbols. The error in the primitive approximation $u_{\rm
         P.A.}$ does not decay to zero, underlining contributions of
         the neutralizing charge background (dot-dashed line).}
\label{FK_fig}
\end{figure}

The existence of the background term $u_{\rm BG}$ can be understood in
two ways. First, by introducing the residual of Bloch equation,
\beq
   R(\rr,\rr';\tau) \equiv \frac{1}{\rho(\rr,\rr';\tau)} \left[ \frac{\partial }{\partial \tau} -
   \lambda_{ij} \, \nabla^2_\rr \,+\, V \right] \rho(\rr,\rr';\tau) 
\label{res}
\;.
\eeq
The residual derived from the primitive action leads to the following
simple expression on the diagonal,
\beq
R(\rr,\rr;\tau) = \frac{\lambda\tau}{2} \nabla^2_\rr V  
                - \frac{\lambda \tau^2 }{4} (\nabla_\rr V)^2
\;\;.
\eeq
For the Ewald potential, the presence of a neutralizing charge
background ($\nabla^2_\rr V_\EW = 4 \pi / \Omega$) gives rise a
constant that is independent of $\rr$. An additional correction for it
can be derived assuming a locally harmonic potential~\cite{Fe72} at
the point $\rr^*=(L/2,L/2,L/2)$,
\beq
u_{\rm BG} = \frac{2}{3} Q_i Q_j \frac{\pi \lambda \tau^2}{\Omega}
-\sum_\LL \Delta u(\rr^*+\LL,\rr^*+\LL;\tau)
\;\;.
\eeq
The first term is of order $O(\tau^2)$ and is therefore not part of
the primitive approximation. A constant shift in the action does not
affect PIMC sampling, however, the $\tau$ derivative of $u_{\rm BG}$
enters in the estimator for kinetic energy. The second way to derive
the background term is by numerically calculations using the
Feynman-Kac formula~\ref{FK}. In Fig.~\ref{FK_fig}, the difference of
the various approximations for the Ewald action are compared with the
computed exact values. The graphs shows that the error in the
primitive approximation approaches a constant at large $r$ while it
decays to zero for a pair of isolated particles. The difference
primarily arises from the presence of the neutralizing background,
which was corrected by introducing $u_{\rm BG}$. However, it should be
stressed that, $\tau=1$ in Fig.~\ref{FK_fig} is much larger than
typical time steps used in simulations with electrons and nuclei. In
such simulations, the background contributions are less importance
because they scale like $\tau^2$.

In analogy to Ewald approach (Eq.~\ref{v_tot2}), the action of
a many-body system can be written as,
%\beqs
%U(\RR,\RR;\tau) &=& \sum_{i>j} \, {u_\EW}_{ij}(\rr_{ij},\rr'_{ij};\tau)
%\; + \;
%\sum_i {u_{\rm M}}_{i}
%\;,\\
%\label{u_tot2}
%{u_{\rm M}}_{i}   &=& \tau \, Q_i^2 V_{\rm M} + 
%                      \frac{1}{2} \sum_{\LL \ne 0} \left[ \bar{u}_{ii}(\LL,0) - \tau \, Q_i^2 \, V_c(\LL)\right]
%\eeqs
%
\beq
U(\RR,\RR';\tau) = \sum_{i>j} \, {u_\EW}_{ij}(\rr_{ij},\rr'_{ij};\tau)
\; + \;
\sum_i \tau \, Q_i^2 V_{\rm M}
\;.
\label{u_tot2}
\eeq
where the charge factors have been included in the action terms. The
efficient evaluation of ${u_\EW}$ is discussed in
Appendix~\ref{section_ewald}.

\section{Results}

Tables~\ref{pe-r-space} and \ref{ee-r-space} give the density matrices
and their $\tau$ derivative for isolated proton-electron and
electron-electron pairs at a time step of 0.125 Hartrees$^{-1}$, which is
typical for a PIMC simulation. In the notation of Eq.~\ref{offd-s},
the diagonal and first order off-diagonal expansion coefficients are
tabulated.

\begin{table}[p]
\caption{ Density matrix at $\tau^{-1}=8$ for an isolated proton-electron pair. The 
	 columns correspond to diagonal potential action, the first
	 off-diagonal expansion coefficient (see Eq.~\ref{offd-s}
	 with $n=1$), the diagonal $\tau$ derivative, and its first
	 expansion coefficient.}  {%\tiny
\vspace*{2mm}
\label{pe-r-space}
\begin{tabular}{r || r|r || r|r}
r & $u(r,0)$  & $A(r)$  
  & $u'(r,0)$ & $A'(r)$ \\
\hline
0.0  & --9.052629e--01   & --1.0656e+00    & --3.701269e+00   & ~4.6908e+00     \\
0.1  & --7.094574e--01   & --7.8668e--01   & --3.664297e+00   & ~2.2366e+00     \\
0.2  & --5.375241e--01   & --5.4357e--01   & --3.435818e+00   & ~4.3544e--01    \\
0.3  & --4.052122e--01   & --3.5276e--01   & --2.988023e+00   & --5.3328e--01   \\
0.4  & --3.130127e--01   & --2.1872e--01   & --2.468529e+00   & --8.1126e--01   \\
0.5  & --2.512131e--01   & --1.3263e--01   & --2.019987e+00   & --7.2479e--01   \\
0.6  & --2.090487e--01   & --8.0229e--02   & --1.683282e+00   & --5.3529e--01   \\
0.7  & --1.789567e--01   & --4.6185e--02   & --1.438155e+00   & --3.5572e--01   \\
0.8  & --1.564692e--01   & --2.7139e--02   & --1.255464e+00   & --2.1970e--01   \\
0.9  & --1.390225e--01   & --1.7645e--02   & --1.114415e+00   & --1.4285e--01   \\
1.0  & --1.250863e--01   & --1.2256e--02   & --1.002118e+00   & --9.8852e--02   \\
1.1  & --1.136946e--01   & --8.9124e--03   & --9.105148e--01  & --7.1705e--02   \\
1.2  & --1.042075e--01   & --6.7064e--03   & --8.343272e--01  & --5.3873e--02   \\
1.3  & --9.618330e--02   & --5.1838e--03   & --7.699461e--01  & --4.1601e--02   \\
1.4  & --8.930751e--02   & --4.0954e--03   & --7.148140e--01  & --3.2843e--02   \\
1.5  & --8.334981e--02   & --3.2949e--03   & --6.670655e--01  & --2.6411e--02   \\
1.6  & --7.813769e--02   & --2.6922e--03   & --6.253068e--01  & --2.1571e--02   \\
1.7  & --7.353934e--02   & --2.2291e--03   & --5.884752e--01  & --1.7856e--02   \\
1.8  & --6.945233e--02   & --1.8672e--03   & --5.557459e--01  & --1.4954e--02   \\
1.9  & --6.579581e--02   & --1.5801e--03   & --5.264687e--01  & --1.2653e--02   \\
2.0  & --6.250516e--02   & --1.3493e--03   & --5.001243e--01  & --1.0803e--02   \\
2.1  & --5.952805e--02   & --1.1616e--03   & --4.762926e--01  & --9.2992e--03   \\
2.2  & --5.682170e--02   & --1.0073e--03   & --4.546301e--01  & --8.0633e--03   \\
2.3  & --5.435076e--02   & --8.7928e--04   & --4.348534e--01  & --7.0381e--03   \\
2.4  & --5.208581e--02   & --7.7215e--04   & --4.167263e--01  & --6.1803e--03   \\
2.5  & --5.000210e--02   & --6.8181e--04   & --4.000506e--01  & --5.4569e--03   \\
2.6  & --4.807872e--02   & --6.0507e--04   & --3.846586e--01  & --4.8426e--03   \\
2.7  & --4.629784e--02   & --5.3947e--04   & --3.704075e--01  & --4.3174e--03   \\
2.8  & --4.464419e--02   & --4.8304e--04   & --3.571749e--01  & --3.8657e--03   \\
2.9  & --4.310460e--02   & --4.3425e--04   & --3.448554e--01  & --3.4752e--03   \\
3.0  & --4.166768e--02   & --3.9181e--04   & --3.333576e--01  & --3.1355e--03   \\
\end{tabular}                          
}
\end{table}

\begin{table}[p]
\caption{ Pair density matrix for $\tau^{-1}=8$ for isolated pair of electrons in the format of Tab.~\ref{pe-r-space}.}
\label{ee-r-space}
\vspace*{2mm}
{%\tiny
\begin{tabular}{r||r|r||r|r}
r & $u(r,0)$  & $A(r)$  
  & $u'(r,0)$ & $A'(r)$ \\
\hline
0.0  & 6.176418e--01   & 4.4659e--01   &  2.435490e+00     & --1.6593e+00   \\
0.1  & 5.191880e--01   & 3.3730e--01   &  2.424522e+00     & --9.0614e--01   \\
0.2  & 4.290318e--01   & 2.4933e--01   &  2.358555e+00     & --3.6651e--01   \\
0.3  & 3.522642e--01   & 1.8106e--01   &  2.218501e+00     & --2.2177e--02   \\
0.4  & 2.903868e--01   & 1.2962e--01   &  2.021059e+00     & ~1.6609e--01     \\
0.5  & 2.422795e--01   & 9.1878e--02   &  1.799564e+00     & ~2.4375e--01     \\
0.6  & 2.055164e--01   & 6.4773e--02   &  1.584616e+00     & ~2.5324e--01     \\
0.7  & 1.774303e--01   & 4.5605e--02   &  1.394356e+00     & ~2.2765e--01     \\
0.8  & 1.557096e--01   & 3.2174e--02   &  1.234499e+00     & ~1.8896e--01     \\
0.9  & 1.385903e--01   & 2.2809e--02   &  1.103098e+00     & ~1.4953e--01     \\
1.0  & 1.248157e--01   & 1.5221e--02   &  9.952834e--01    & ~1.1214e--01     \\
1.1  & 1.135143e--01   & 1.0485e--02   &  9.060434e--01    & ~8.1349e--02     \\
1.2  & 1.040821e--01   & 7.6096e--03   &  8.312447e--01    & ~6.0020e--02     \\
1.3  & 9.609319e--02   & 5.7381e--03   &  7.677422e--01    & ~4.5517e--02     \\
1.4  & 8.924103e--02   & 4.4539e--03   &  7.131934e--01    & ~3.5419e--02     \\
1.5  & 8.329967e--02   & 3.5366e--03   &  6.658461e--01    & ~2.8165e--02     \\
1.6  & 7.809915e--02   & 2.8606e--03   &  6.243711e--01    & ~2.2803e--02     \\
1.7  & 7.350923e--02   & 2.3497e--03   &  5.877450e--01    & ~1.8744e--02     \\
1.8  & 6.942845e--02   & 1.9556e--03   &  5.551675e--01    & ~1.5608e--02     \\
1.9  & 6.577663e--02   & 1.6462e--03   &  5.260046e--01    & ~1.3144e--02     \\
2.0  & 6.248957e--02   & 1.3996e--03   &  4.997475e--01    & ~1.1178e--02     \\
2.1  & 5.951525e--02   & 1.2005e--03   &  4.759834e--01    & ~9.5902e--03     \\
2.2  & 5.681109e--02   & 1.0377e--03   &  4.543740e--01    & ~8.2918e--03     \\
2.3  & 5.434190e--02   & 9.0340e--04   &  4.346395e--01    & ~7.2197e--03     \\
2.4  & 5.207834e--02   & 7.9148e--04   &  4.165462e--01    & ~6.3261e--03     \\
2.5  & 4.999577e--02   & 6.9745e--04   &  3.998978e--01    & ~5.5751e--03     \\
2.6  & 4.807331e--02   & 6.1784e--04   &  3.845282e--01    & ~4.9393e--03     \\
2.7  & 4.629319e--02   & 5.4998e--04   &  3.702955e--01    & ~4.3971e--03     \\
2.8  & 4.464018e--02   & 4.9176e--04   &  3.570782e--01    & ~3.9319e--03     \\
2.9  & 4.310111e--02   & 4.4153e--04   &  3.447714e--01    & ~3.5305e--03     \\
3.0  & 4.166463e--02   & 3.9794e--04   &  3.332844e--01    & ~3.1821e--03     \\
\end{tabular}
}
\end{table}

As a first test of the isolated pair action, a single hydrogen atom at
a temperature $T=0.025$. This temperature was chosen low enough for
the atom to be in the ground state since the relative occupation
probability of the first excited state is $5 \times 10^{-5}$. Path
integral Monte Carlo estimates of the potential energy agrees with the
exact value of $-1$ within the statistical error bar of the Monte
Carlo integration, here $10^{-4}$. While the evaluation of the
potential energy involves only the action, the total energy estimator
requires also the $\tau$ derivative. The computed total energy agreed
with the exact value of $-1/2$ to a relative accuracy of $7 \times
10^{-4}$. Both potential and kinetic energy estimators are necessary
to calculate the pressure, $P$, from the virial theorem,
$3\left<P\right>\Omega = 2\left<K\right>+\left<V\right>$ .

\begin{figure}[htb]
\centerline{\includegraphics[angle=0,width=0.85\textwidth]{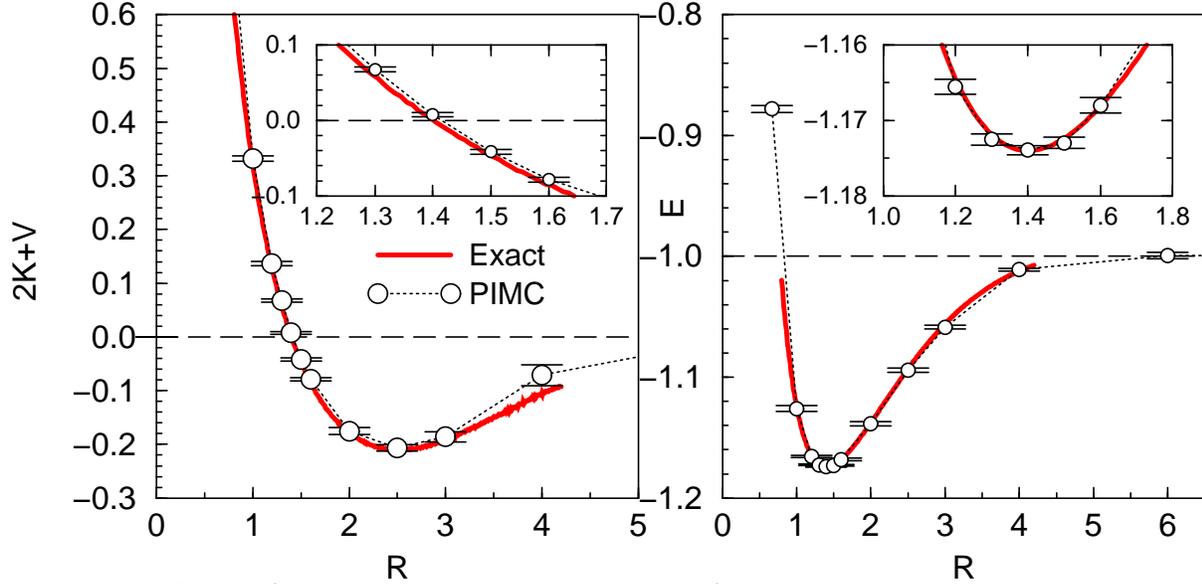}}
\caption{Virial term $2K+V$ (kinetic energy $K$ and potential energy $V$) 
	and total internal energy $E$ are shown for an isolated
	hydrogen molecule, for which the nuclei have been kept fixed at
	separation $R$. The PIMC results have been obtained from
	simulations at $T=0.0125$ with 320 time slices. The solid
	lines are the exact groundstate results of~\cite{KW64}.}
\label{H2_vs_R}
\end{figure}

As a second test for the accuracy of the computed pair density matrix,
the kinetic and potential energy for an isolated hydrogen molecule was
computed as a function of proton separation $R$. Fig.~\ref{H2_vs_R}
shows very good agreement with the exact groundstate
calculation~\cite{KW64}. The molecular binding energy of $1.174448$ as
well as the location of the minimum at $R=1.4008$ are well reproduced
(see inset of figure~\ref{H2_vs_R}).
The virial theorem for a diatomic molecule~\cite{KW64,St76} reads,
\beq
2K+V = - R \, \frac{dE}{dR}\;.
\eeq
It goes to zero at the equilibrium bond length as well as in the limit
of $R \to \infty$ as shown in the figure. 

\begin{figure}[htb]
\centerline{\includegraphics[angle=0,width=0.5\textwidth]{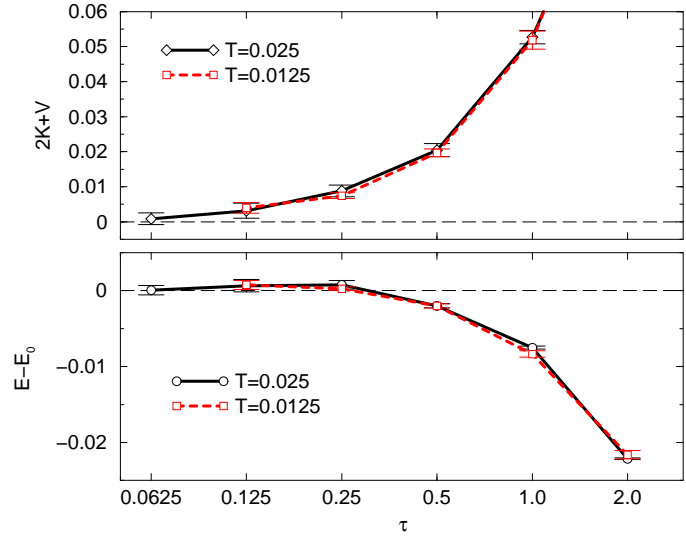}}
\caption{Total internal energy $E$ and virial term $2K+V$ for an isolated
	hydrogen molecule are shown for different temperatures as a
	function of the time step in the path integral. The nuclei
	have been kept fixed at the equilibrium separation of $R=1.4008$. }
\label{H2_acc}
\end{figure}

A more detailed analysis of the comparison with the groundstate
calculation for $R$ fixed at the equilibrium bond length shown in
Fig.~\ref{H2_acc} reveals that the finite time step in the path
integral determines the accuracy of the calculation. The temperature
$T=0.025$ was chosen low enough to compare with the groundstate result
as the check with $T=0.0125$ shows. The internal energy converges
faster to the exact groundstate result than the virial term
$2K+V$. For $\tau \le 0.25$, we find that the energy deviates less
than 10$^{-3}$ from the exact groundstate energy. For $\tau=0.0625$,
the energy deviated by $(2.4~\pm~1.7) \times 10^{-4}$.  For this time
step, we obtained $2K+V = (9 \pm 7) \times 10^{-4}$, which corresponds
to an residual inaccuracy in the pressure equivalent to the pressure
of an ideal molecular gas at $T=90\pm 70\,$K, which is a significant
improvement in accuracy over~\cite{MC01} and more than sufficient for
the majority of PIMC simulations of hot, dense
hydrogen~\cite{MC00,Mi01}.  This concludes the accuracy analysis for
isolated systems of particles.

The accuracy of the constructed pair density matrix
in a periodic system is analyzed by calculating the error in Bloch
equation~\ref{Bloch_eq} for the Ewald potential. 
The optimized Ewald break-up of Eqn.~\ref{break-up} 
for  a cubic simulation cell of size $L=5$ using 20 $k$-shells
is given in Tables~\ref{R-space} and \ref{k-space}. 

\begin{table}
\caption{Real space part $W$ of the action for proton-electron and 
         electron-electron pair density matrices computed for 
         $\tau^{-1}=8$ using an optimized
       Ewald break-up for a cubic simulation cell of size $L=5$. The
       $k$-space part is given in Tab.~\ref{k-space}.}  {%\tiny
\vspace*{2mm}
\label{R-space}
\begin{tabular}{r||r|r||r|r}
r & $W_{\rm pe}(r)$  & $W'_{\rm pe}(r)$ & $W_{\rm ee}(r)$  & $W'_{\rm ee}(r)$\\
\hline
0.0  &  --3.948972e--01   & --2.462608e+00  & ~3.779454e--01 & ~1.247426e+00   \\
0.1  &  --2.350750e--01   & --2.258029e+00  & ~2.919260e--01 & ~1.196514e+00   \\
0.2  &  --1.160451e--01   & --1.852593e+00  & ~2.190901e--01 & ~1.083259e+00   \\
0.3  &  --3.789284e--02   & --1.376056e+00  & ~1.608496e--01 & ~9.211309e--01  \\
0.4  &  ~8.202359e--03    & --9.633273e--01 & ~1.163420e--01 & ~7.390493e--01  \\
0.5  &  ~3.255803e--02    & --6.601111e--01 & ~8.329489e--02 & ~5.647894e--01  \\
0.6  &  ~4.338204e--02    & --4.511075e--01 & ~5.925357e--02 & ~4.155765e--01  \\
0.7  &  ~4.616727e--02    & --3.082067e--01 & ~4.188614e--02 & ~2.972913e--01  \\
0.8  &  ~4.420917e--02    & --2.096609e--01 & ~2.939746e--02 & ~2.081889e--01  \\
0.9  &  ~3.948008e--02    & --1.413132e--01 & ~2.041829e--02 & ~1.432097e--01  \\
1.0  &  ~3.338766e--02    & --9.395882e--02 & ~1.400554e--02 & ~9.683606e--02  \\
1.1  &  ~2.703458e--02    & --6.140030e--02 & ~9.466156e--03 & ~6.429422e--02  \\
1.2  &  ~2.110379e--02    & --3.929014e--02 & ~6.284010e--03 & ~4.181878e--02  \\
1.3  &  ~1.587569e--02    & --2.453568e--02 & ~4.085855e--03 & ~2.657455e--02  \\
1.4  &  ~1.142789e--02    & --1.489252e--02 & ~2.581234e--03 & ~1.644121e--02  \\
1.5  &  ~7.866995e--03    & --8.749983e--03 & ~1.587436e--03 & ~9.870341e--03  \\
1.6  &  ~5.231115e--03    & --4.948775e--03 & ~9.427062e--04 & ~5.723989e--03  \\
1.7  &  ~3.356079e--03    & --2.677517e--03 & ~5.456568e--04 & ~3.186790e--03  \\
1.8  &  ~2.013496e--03    & --1.378287e--03 & ~3.001534e--04 & ~1.692584e--03  \\
1.9  &  ~1.098643e--03    & --6.634200e--04 & ~1.574300e--04 & ~8.477546e--04  \\
2.0  &  ~5.739410e--04    & --2.995616e--04 & ~7.289096e--05 & ~3.991008e--04  \\
2.1  &  ~3.066066e--04    & --1.223001e--04 & ~2.820100e--05 & ~1.695370e--04  \\
2.2  &  ~1.462296e--04    & --4.510938e--05 & ~1.383209e--05 & ~6.587290e--05  \\
2.3  &  ~3.113736e--05    & --1.503693e--05 & ~3.426564e--06 & ~1.986137e--05  \\
2.4  &  --1.417605e--05   & --4.564337e--06 & ~6.935208e--07 & ~6.273595e--06  \\
2.5  &  0                 & 0               & 0              &  0              \\
\end{tabular}
}
\end{table}

\begin{table}
\caption{K-space action for proton-electron and electron-electron pair
       density matrices computed for $\tau^{-1}=8$ using an optimized
       Ewald break-up with 20 k-shells for a cubic simulation cell of
       size $L=5$. The corresponding real space parts are listed in
       Tab.~\ref{R-space}. The table also includes the Madelung and
       the background constant from~Eqs.~\ref{u_tot2}
       and~\ref{break-up}.}
\label{k-space}
{%\tiny
\vspace*{2mm}
\begin{center}
\begin{tabular}{c||r|r||r|r}
$n^2, k={2 \pi n}/{L}$ & $\tilde{u}_{k,\rm pe}$ & $\tilde{u}'_{k,\rm pe}$ & $\tilde
{u}_{k,\rm ee}$ & $\tilde{u}'_{k,\rm ee}$  \\
\hline
$u_{\rm M}$ & ~3.5466e--02  & ~2.8372e--01  & --3.5467e--02   & 2.8376e--01  \\
$C_u$       & --3.0502e--03   & ~2.0762e--02   &  --2.9417e--03  & --2.2041e--02 \\
\hline
1    & --9.8752e--03  & --4.5714e--02  & ~5.4742e--03   & ~4.4979e--02 \\
2    & --5.0783e--03  & --1.6276e--02  & ~1.8624e--03   & ~1.5781e--02 \\
3    & --3.1684e--03  & --7.6607e--03  & ~8.3439e--04   & ~7.3217e--03 \\
4    & --2.0962e--03  & --4.0198e--03  & ~4.1473e--04   & ~3.7862e--03 \\
5    & --1.4182e--03  & --2.2280e--03  & ~2.1645e--04   & ~2.0676e--03 \\
6    & --9.6702e--04  & --1.2727e--03  & ~1.1557e--04   & ~1.1634e--03 \\
8    & --4.4850e--04  & --4.3268e--04  & ~3.3300e--05   & ~3.8330e--04 \\
9    & --3.0283e--04  & --2.5364e--04  & ~1.7587e--05   & ~2.2113e--04 \\
10   & --2.0262e--04  & --1.4824e--04  & ~9.0566e--06   & ~1.2710e--04 \\
11   & --1.3405e--04  & --8.5951e--05  & ~4.4823e--06   & ~7.2484e--05 \\
12   & --8.7505e--05  & --4.9194e--05  & ~2.0841e--06   & ~4.0853e--05 \\
13   & --5.6226e--05  & --2.7704e--05  & ~8.7102e--07   & ~2.2662e--05 \\
14   & --3.5469e--05  & --1.5293e--05  & ~2.9069e--07   & ~1.2311e--05 \\
16   & --1.3175e--05  & --4.2131e--06  & --5.3452e--08  & ~3.3414e--06 \\
17   & --7.6876e--06  & --2.0560e--06  & --7.1360e--08  & ~1.6470e--06 \\
18   & --4.3189e--06  & --9.4808e--07  & --6.0744e--08  & ~7.6995e--07 \\
19   & --2.3129e--06  & --4.0035e--07  & --4.2831e--08  & ~3.3463e--07 \\
20   & --1.1631e--06  & --1.3415e--07  & --2.6457e--08  & ~1.3167e--07 \\
21   & --5.3551e--07  & --1.5653e--08  & --1.4397e--08  & ~4.4726e--08 \\
22   & --2.1498e--07  & ~1.9113e--08   & --6.7088e--09  & ~1.1253e--08 \\
\end{tabular}
\end{center}
}
\end{table}

\begin{figure}[htb]
\centerline{\includegraphics[angle=0,width=0.5\textwidth]{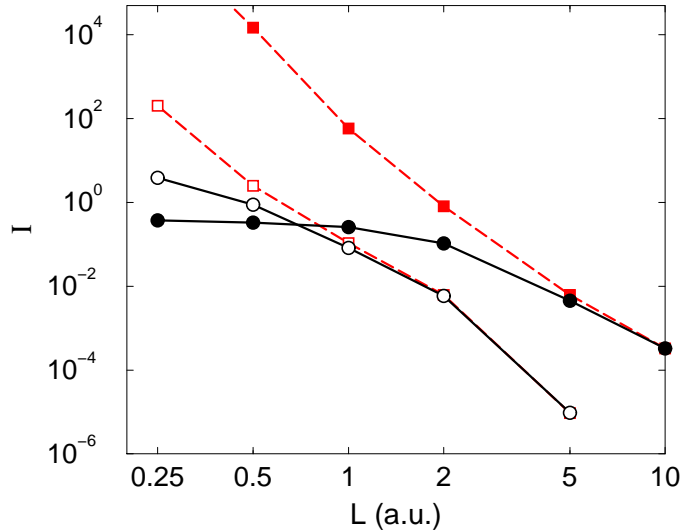}}
\caption{Residual of the Bloch equation $I(L;\tau)$ in periodic boundary conditions
         is shown as function of cell size $L$ for the proton-electron
         pair density matrix at two different temperatures (open
         symbols $\tau^{-1}=8$ and full symbols $\tau^{-1}=0.5$). The
         pair approximation for periodic images (circle) is compared
         with treating the images in the primitive approximation
         (squares).
}
\label{residual}
\end{figure}

Based on the residual in Eq.~\ref{res}, we define the accuracy parameter,
\beq
\left.
I(L;\tau) = \int \!\!\! \int \dd\rr \, \dd\rr' \rho(\rr,\rr';\tau) 
\left| R(\rr,\rr';\tau) \right|
\right / 
\int\!\!\!\int \dd\rr \, \dd\rr' \rho(\rr,\rr';\tau) \;,
\eeq
that characterizes the accuracy of the particular approximation to the
density matrix. Fig.~\ref{residual} shows this accuracy parameter for
the periodic proton-electron pair density matrix as a function of cell
size $L$. This present method of treating the image charges in the
pair approximation is contrasted with treating them in the primitive
approximation, Eqn.~\ref{prim-approx}, with the long range term
$V_{l.r.}({\bf r})=V_{EW}({\bf r})-1/r$. In the limit of large $L$,
both methods are in agreement and show a rapid decay of the residual
indicating that the Bloch equation is satisfied with increasing
accuracy. For small $L$, the pair approximation yields a smaller
residual demonstrating that the method proposed here leads to a more
accurate high temperature density matrix for periodic systems.

\begin{figure}[htb]
\centerline{\includegraphics[angle=0,width=0.85\textwidth]{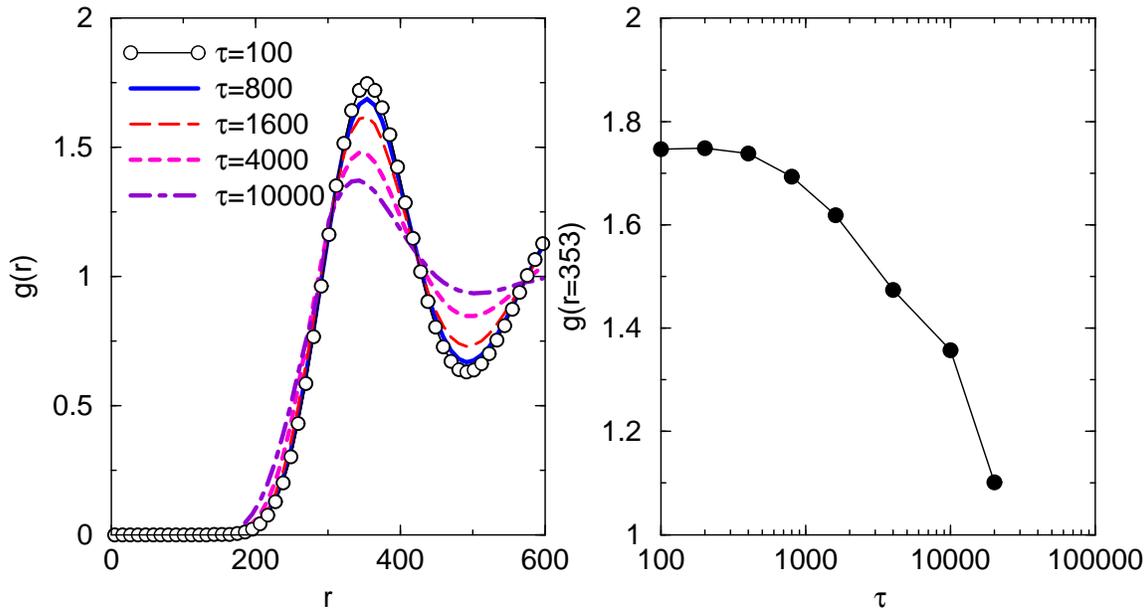}}
\caption{The pair correlation function $g(r)$ for the homogeneous electron gas
	in state of a Wigner crystal~\cite{Jo96,MP04,MP05,MG06} for
	coupling parameter $\Gamma=200$ and quantumness parameter
	$\eta=1$ ($\beta=40000, r_s=200$) are shown from simulation
	with different time steps $\tau$. The right graph shows how
	the maximum value of the $g(r)$ converges as a function of
	$\tau$. Good convergence is reached for $\tau=400$ requiring
	simulations with 100 time slices. }
\label{gr}
\end{figure}

We have applied Eq.~\ref{u_EW} to PIMC simulations of hydrogen,
helium, hydrogen-helium
mixtures~\cite{Mi05,Mi06,SCCS2008Militzer,Mi09}, and to study quantum
effects in the one-component
plasma~\cite{MP04,MP05,MG06}. Fig.~\ref{gr} demonstrates the quality
of the construct density matrices in PIMC simulation of a strong
coupled Coulomb system. Despite significant quantum properties the
particles arrange in a Wigner crystal. The presented time step
analysis shows that a time step of 400 or less is needed to obtain
converged structural properties such as pair correlation
functions. This serves as a benchmark for future methods to construct
density matrices for PIMC that would then need to demonstrate that the
same level of accuracy can be achieved with larger time steps.

\section{Conclusions}

     The method proposed here to compute the Coulomb pair density matrix in
periodic boundary conditions has two advantages over previous
approaches. It starts from the more accurate Coulomb pair density
matrix, which reduces the numerical errors in the short-range behavior
of the computed pair density matrix and circumvents a break-up of the
potential. Although it does not derive the pair density matrix for the
Ewald potential exactly, it consistently uses the pair approximation for 
the action which is an intrinsic assumption in most many-body
path integration simulations.

   In addition to describing this method, 
quantitative test results for the isolated
hydrogen atom and molecule as well as a test for the periodic system 
based on the residual error in the Bloch equation were presented
to assess the accuracy of the high temperature density matrix.
It is suggested that these tests should form a basis for comparing
any proposed alternative methods.

\appendix

\section{Ewald potential}
\label{section_ewald} 

The Ewald method to describe the potential of charged particles in
periodic boundary conditions can be adapted in the following way to
calculate the action. The potential energy of a system of $N$ charges
$Q_i$ interacting via the Coulomb potential $V_c(\rr) \equiv 1/|\rr|$
is given by~\cite{AT87},
\beq
V = \sum_{i>j} \sum_\LL Q_i Q_j \, V_c(\rr_{ij}+\LL)
\; + \;
\frac{1}{2} \sum_{\LL \ne 0} \sum_i Q_i^2 \, V_c(\LL)
\;.
\label{pot2}
\eeq
where $\rr_{ij}=\rr_i-\rr_j$ and $\LL$ is a lattice vector. This
expression converges {\it conditionally} for charge neutral systems
($\sum_i Q_i=0$). 
Under the assumption of charge neutrality, this expression can be
computed from Ewald potential $V_\EW$ and the Madelung constant $V_{\rm M}$,
\beq
V = \sum_{i>j} Q_i Q_j \, V_\EW(\rr_{ij})
\; + \;
\sum_i Q_i^2 \, V_{\rm M}
\label{v_tot2}
\;,
\eeq
where $V_{\rm M} = \frac{1}{2} \lim_{r \to 0} [ V_\EW(r) - V_c(r) ].$
The Ewald potential is split in a real space and a Fourier
part~\cite{Ew17} by subtracting the screening potential $V_s(r)$ of
a Gaussian charge distribution. Under the
assumption of charge neutrality ($\sum_i Q_i = 0$) or in the presence
of a neutralizing background, the Ewald potential becomes,
\beq
V_\EW(\rr) = \sum_{\LL} \left[V_c-V_s\right](\rr+\LL) 
	     + \sum_{\kk \ne 0} \tilde{V}_s(\kk) e^{i\kk \cdot \rr} + C_V \;,
\eeq
where the constant $C_V$,
\beq
C_V = - \frac{1}{\Omega} \int \! \dd^3\rr\, \left[ V_c-V_s \right] (\rr) \;,
\eeq
represents contributions from a neutralizing background. For neutral
systems, all $C_V$ terms cancel. 

The total potential energy can be computed more efficiently by
introducing the Fourier transform of the charge density, $\tilde \rho(\kk) =
\sum_j \, Q_j \, e^{-i \kk \rr_j}.$ Eq.~\ref{pot2} then reads,
\beq
V = \sum_{i>j} Q_i Q_j \, \left[ \sum_{\LL} \left[V_c-V_s\right](\rr+\LL) + 
\frac{1}{2}  C_V \, |\tilde\rho(0)|^2 + \frac{1}{2} \sum_{\kk \ne 0} \tilde{V}_s(\kk) \, |\tilde\rho(\kk)|^2 \right] 
+ \sum_i \, Q_i^2 \, V_{\rm image} \;, 
\eeq
where the constant $V_{\rm image}$ has been introduced,
\beq
2 V_{\rm image} = 2 V_{\rm M} - C_V - \sum_{\kk \ne 0} \tilde{V}_s(\kk) 
= \sum_{\LL \ne 0} \left[V_c-V_s\right](\LL) - V_s[0] \;.
\eeq
In contrast to $V_{\rm M}$, $V_{\rm image}$ depends on the choice of
$V_s$.

The same approach can be applied to construct the diagonal Ewald
action where $V_c(r)$ is then replaced by $\tau V_c(r) + \Delta
u(\rr,\rr;\tau)$. In many-body Monte Carlo simulations, an efficient
computation of the Ewald action, Eq.~\ref{u_EW}, is crucial. This can
be achieved by using the optimized Ewald break-up technique developed
by Natoli {\em et al.}~\cite{Na95},
\beq
u_\EW(\rr) \approx \sum_n a_n f_n(|\rr|) + \sum_{|\kk| \le k_c}{y_{|\kk|} \, e^{-i \kk \cdot \rr}} + C_u \;.
\label{break-up}
\eeq
where one uses only one real-space image and a variable number of
Fourier vectors depending on the required accuracy. As basis set
$f_n$, we use locally piecewise quintic Hermite interpolants as
suggested in~\cite{Na95}. However, this basis set is only used to
perform the break-up.

The coefficients $a_n$ and $y_\kk$ are chosen to minimize the mean
squared deviation between the $u_\EW(\rr)$ and the fit function. While
Natoli {\em et al.} proposed using a sum over many Fourier components
for this step, we reached a higher accuracy by performing this
calculation in real space, which leads to the usual set of $n$ linear
equations,
\beq
0 = u_{{\rm EW},n} - \sum_{|\kk| \le k_c} \tilde{u}_\EW(\kk) \tilde{f}_n(\kk) 
-\sum_m a_m \left[ f_{m,n} - \sum_{|\kk| \le k_c} \tilde{f}_m(\kk) \tilde{f}_n(\kk) \right]
\eeq
with
\beq
u_{{\rm EW},n} = \frac{1}{\Omega} \int \! d^3\rr \, u_{\rm EW}(\rr) \, f_m(\rr) 
\;\;\;\;, 
\;\;\;\; 
f_{n,m} = \frac{1}{\Omega} \int \! d^3\rr \, f_n(\rr) \, f_m(\rr) 
\;\;\;\;,
\eeq
and $\tilde{f}_n$ and $\tilde{u}_\EW$ being the corresponding Fourier
transforms of $f_n$ and $u_\EW$.
The Fourier coefficients $y_\kk$ are given by,
\beq
y_\kk = \tilde{u}_\EW(\kk) - \sum_n \: a_n \: \tilde{f}_n(\kk) \;.
\eeq
The background term, $C_u$, is treated as a $k=0$ component. In the
final step, the real space part,
\beq
W(|\rr|)=\sum_n a_n f_n(|\rr|)\;\;,
\label{w}
\eeq
is conveniently stored on a radial grid and interpolated during PIMC
simulations. The same procedure is applied to the $\tau$ derivative of
the action.

\section*{Acknowledgments}

E.L. Pollock contributed several important ideas to this
manuscript. We also thank Ken Esler, Bernard Bernu, and David Ceperley
for useful discussions. We acknowledge support from NSF and NASA.

%\bibliographystyle{unsrt}
%\bibliography{pimc}

\end{document}